**PCCP**

## ARTICLE

# Structure and crystallization of SiO₂ and B₂O₃ doped lithium disilicate glasses from theory and experiment

Andreas Erlebach, Katrin Thieme,[a] Marek Sierka,[b] Christian Rüssel

Solid solutions of SiO₂ and B₂O₃ in Li₂O·2SiO₂ are synthesized and characterized for the first time. Their structure and crystallization mechanisms are investigated employing a combination of simulations at the density functional theory level and experiments on the crystallization of SiO₂ and B₂O₃ doped lithium disilicate glasses. The remarkable agreement of calculated and experimentally determined cell parameters reveals the preferential, kinetically controlled incorporation of [SiO₄] and [BO₄] at the Li⁺ lattice sites of the Li₂O·2SiO₂ crystal structure. While the addition of SiO₂ increases the glass viscosity resulting in lower crystal growth velocities, glasses containing B₂O₃ show a reduction of both viscosities and crystal growth velocities. These observations could be rationalized by a change of the chemical composition of the glass matrix surrounding the precipitated crystal phase during the course of crystallization, which leads to a deceleration of the attachment of building units required for further crystal growth at the liquid-crystal interface.

## 1 Introduction

Metastable crystal phases and solid solutions can have superior materials properties compared to their thermodynamically more stable counterparts,[1] e.g., an enhanced ion conductivity[2] or an improved photocatalytic activity.[3] However, metastable crystals including those that take up dopants in their crystal lattice can only be formed or 'trapped' by kinetically controlled processes. Therefore, the elucidation of the underlying formation mechanisms – nucleation and crystal growth – is essential for the targeted crystallization of the desired phase starting from a solution[4,5] or an amorphous state.[6,7] Accordingly, a profound understanding of these crystallization processes plays a crucial role in the optimization of synthesis routes of materials with tailored properties. For instance, the properties of crystalline lithium silicates and, consequently, their fields of applications ranging from Li battery materials[8] through tritium breeder materials in fusion reactors[9] to glass-ceramics in dentistry,[10] strongly depend on the compounds and phases formed during the crystallization.

Most studies on the crystallization of glass employed models for isochemical systems, i.e., systems in which the chemical composition is equal to the composition of the precipitated crystal phase.[11–15] In these systems, the chemical composition does not change during the course of the crystallization. By contrast, in non-isochemical systems, the composition and hence also the physical properties of the residual melt will change with time.[16] This was frequently studied in the past few years especially for the crystallization of alkaline earth or rare earth fluorides from oxyfluoride glasses[17–19] and multicomponent lithium disilicate glasses.[20–23] In that case components of the glass which decrease the melt viscosity are removed leading to an increase in viscosity near the formed crystals.[24] This results in formation of a highly viscous shell around the growing particles, in particular when the crystallization takes place at temperatures with relatively high glass viscosities.[17,25] After the viscosity of the shell exceeds a value of 10¹³ dPa·s, i.e. the glass transition temperature is above the temperature supplied during crystallization, crystal growth is completely stopped.[24] These core shell structures have been experimentally evidenced by anomalous X-ray scattering (ASAXS)[26,27] as well as by high resolution transmission electron microscopy (TEM) especially using electron energy loss spectroscopy (EELS).[28,29] It has been shown that similar mechanisms might occur also in silicate systems.[30] Tailoring the glass composition for this purpose is a versatile tool for the preparation of glass-ceramics with a high volume concentration of nanometer sized crystals.

In any non-isochemical system in which one melt component is not incorporated into the formed crystalline phase and hence is shoved away during crystal growth, concentration gradients are formed, which affect further crystal growth and a time dependent crystal growth velocity might be obtained. By contrast, a competing effect might be that the component shoved away in front of the growing crystal decreases the viscosity and hence might increase the diffusivities.[31] Then the crystal growth velocity might increase as a function of time.

Otto-Schott-Institut für Materialforschung (OSIM), Friedrich-Schiller-Universität Jena, 07743 Jena, Germany.
ᵃ Katrin Thieme, E-mail: katrin.thieme@uni-jena.de.
ᵇ Marek Sierka, E-mail: marek.sierka@uni-jena.de.

The isochemical glass system most frequently studied in the past was the $Li_2O \cdot 2SiO_2$ system (*e.g.*, Refs. 32–39). Here, nucleation rates as well as crystal growth velocities were studied in the whole temperature range, from the glass transition temperature to the melting temperature.[40] Recently, also the effects of some additives, such as $ZrO_2$,[41] $La_2O_3$, $TiO_2$, $Al_2O_3$,[42] $CeO_2$, $Y_2O_3$,[43] $Nb_2O_5$ and $Ta_2O_5$[44] were reported. Moreover, it was shown that the addition of, *e.g.*, $Al_2O_3$ and MgO improves the mechanical properties of lithium silicate glass ceramics.[45,46] The addition of these components leads to a decrease of the nucleation rate by several orders of magnitude, to a significant increase of the induction period and the deceleration of crystal growth.[42] Moreover, it was shown that some of these compounds are preferably located at the interface crystal/liquid and hence decrease the interfacial energy.[42,43,47] The layer at the interface acts as a barrier for the attachment of new building units required for crystal growth and hence results in an overall decrease of the nucleation rate and crystal growth velocities.[47] Recently, also the effect of additives which should lower the viscosity such as alkalis and $B_2O_3$ were investigated.[48] These components should be enriched in the residual glass matrix leading to a decrease of the viscosity and thus higher crystal growth velocities.

This, however, is only the case if these components are really enriched in the glassy matrix and are not incorporated in the crystal. Consequently, it is quite essential to know whether these compounds are incorporated in the host lattice and, if so, in which quantities. Accordingly, it has to be determined, if there is a thermodynamic driving force for formation of such solid solutions during crystallization. One of the key prerequisites to demonstrate the existence and thermodynamic stability of a crystalline solid solution is the knowledge how dopants are incorporated into the crystal lattice, *i.e.*, which crystallographic lattice site is occupied by substitutional ions. Furthermore, the situation gets more complex if there is a difference between the valence of the dopant and the ions of the host lattice, such as the incorporation of $B^{3+}$ into $Li_2O \cdot 2SiO_2$, due to the required creation of additional lattice defects to preserve charge neutrality. The incorporation of one or more lattice defects and the possible formation of defect cluster strongly influences the crystal structure, its thermodynamic stability and the materials properties, *e.g.*, diffusivity and ion conductivity.[49–51] Therefore, the detailed characterization of the crystal structure at an atomistic level is of fundamental importance. However, obtaining such structural information solely from experiments can be a very challenging task, in particular in case of low dopant concentrations or nanocrystalline materials.

In this context, computational approaches employing simulations at the density functional theory (DFT) level proved to be an efficient tool not only for interpretation of experimental data but also for the prediction of new structures and properties of perfect[52] as well as defect containing crystals.[53] Based on the knowledge of the atomic structure for certain chemical compositions, the thermodynamic stability of these configurations can be evaluated by DFT simulations providing vital information on the incorporation of dopants into the crystal lattice of solid solutions.[54–56]

This work provides computational studies on the incorporation of $Si^{4+}$ and $B^{3+}$ into a $Li_2O \cdot 2SiO_2$ host lattice in combination with experiments on the crystallization of $Li_2O \cdot 2SiO_2$ glasses to which a certain concentration of $B_2O_3$ or $SiO_2$ were added. By revealing the crystal structure of the precipitated phase at an atomic level, it is possible to provide the first clear indication for the existence of metastable $x$ $B_2O_3 \cdot (100-x)$ $((33.3-y)Li_2O \cdot (66.7+y)SiO_2)$ solid solutions. Along with the determination of crystal growth velocities and glass viscosities, conclusions on the underlying glass crystallization processes can be drawn.

## 2 Computational details

Density functional theory (DFT) calculations employed the Vienna *Ab Initio* Simulation Package (VASP)[57,58] along with the projector augmented wave (PAW) method.[59,60] In order to investigate the accuracy of DFT simulations structure optimizations were performed using both the Perdew-Burke-Ernzerhof (PBE)[61,62] exchange-correlation functional and its revised version for solids (PBEsol).[63,64] Best agreement with experiments (*cf.* Computational results) was achieved by employing the PBEsol exchange-correlation functional in combination with the empirical dispersion correction of Grimme *et.al.* (DFT-D3).[65] Therefore, this setup was applied for the remaining calculations. All structure optimizations were performed under constant (zero) pressure conditions along with an energy cutoff of 800 eV for the plane wave basis sets. The integration of the first Brillouin employed Monkhorst-Pack[66] grids with a line density of about 15 *k*-points per $Å^{-1}$ along each reciprocal lattice vector.

To model the solid solution $x$ $B_2O_3 \cdot (100-x)$ $((33.3-y)Li_2O \cdot (66.7+y)SiO_2)$ three supercells with different sizes, denoted as **0a**, **0b** and **0c** were constructed from the optimized crystal structure of $Li_2O \cdot 2SiO_2$. For **0a** and **0b** the vector (**a b c**) containing the basis vectors **a**, **b** and **c** of the $Li_2O \cdot 2SiO_2$ unit cell were multiplied with the matrices $X_{0a}$ and $X_{0b}$

$$X_{0a} = \begin{pmatrix} 1 & 0 & -2 \\ 0 & 1 & 0 \\ 3 & 0 & 1 \end{pmatrix} X_{0b} = \begin{pmatrix} 1 & 0 & -1 \\ 0 & 1 & 0 \\ 1 & 0 & 2 \end{pmatrix}, \qquad (1)$$

respectively. The new basis vectors (**a*** **b*** **c***) are a linear combination of the old ones and are calculated from (**a*** **b*** **c***) = (**a b c**) $X$. In addition, a 1×1×2 supercell of $Li_2O \cdot 2SiO_2$ (**0c**) was constructed.

In order to introduce excess of $SiO_2$ and for incorporation of $B_2O_3$ defects were created in **0a**, **0b** and **0c**. The defects were created at the $Li^+$ (models **1** and **3**) and $Si^{4+}$ (models **2** and **4**) lattice sites. This yields structures with an excess of $SiO_2$ (**1** and **2**, $y > 0$) as well as unit cells containing $B_2O_3$ (**3** and **4**, $x > 0$). The models were generated by combining ion substitutions with vacancies to preserve charge neutrality without making additional assumptions regarding the ionic radii.

**Table 1** Unit cell composition of $x$ B$_2$O$_3$·(100-$x$) (($33.3-y$)Li$_2$O·($66.7+y$)SiO$_2$) along with the corresponding SiO$_2$ and Li$_2$O concentrations and the incorporated lattice defects in Kröger-Vink notation.

| Model | Unit cell | Chemical composition [mol%] | | | | Defects |
|---|---|---|---|---|---|---|
| | | Li$_2$O | SiO$_2$ | B$_2$O$_3$ ($x$) | $y$ | |
| **0a** | 28 (Li$_2$O · 2 SiO$_2$) | 33.33 | 66.67 | 0.00 | 0.00 | - |
| **0b** | 12 (Li$_2$O · 2 SiO$_2$) | 33.33 | 66.67 | 0.00 | 0.00 | - |
| **0c** | 8 (Li$_2$O · 2 SiO$_2$) | 33.33 | 66.67 | 0.00 | 0.00 | - |
| **1a** | 26 (Li$_2$O · 2.19 SiO$_2$) | 31.32 | 68.68 | 0.00 | 2.01 | Si$_{\text{Li}}^{···}$ + 3v$_{\text{Li}}'$ |
| **1b** | 10 (Li$_2$O · 2.50 SiO$_2$) | 28.57 | 71.43 | 0.00 | 4.76 | Si$_{\text{Li}}^{···}$ + 3v$_{\text{Li}}'$ |
| **1c** | 6 (Li$_2$O · 2.83 SiO$_2$) | 26.09 | 73.91 | 0.00 | 7.25 | Si$_{\text{Li}}^{···}$ + 3v$_{\text{Li}}'$ |
| **2a** | 27 (Li$_2$O · 2.07 SiO$_2$) | 32.53 | 67.47 | 0.00 | 0.80 | v$_{\text{O}}^{··}$ + 2v$_{\text{Li}}'$ |
| **2b** | 11 (Li$_2$O · 2.18 SiO$_2$) | 31.43 | 68.57 | 0.00 | 1.91 | v$_{\text{O}}^{··}$ + 2v$_{\text{Li}}'$ |
| **2c** | 7 (Li$_2$O · 2.29 SiO$_2$) | 30.43 | 69.57 | 0.00 | 2.90 | v$_{\text{O}}^{··}$ + 2v$_{\text{Li}}'$ |
| **3a** | 0.5 (B$_2$O$_3$)· 26.5 (Li$_2$O · 2.11 SiO$_2$) | 31.53 | 66.62 | 1.85 | 1.21 | B$_{\text{Li}}^{··}$ + 2v$_{\text{Li}}'$ |
| **3b** | 0.5 (B$_2$O$_3$)· 10.5 (Li$_2$O · 2.29 SiO$_2$) | 29.05 | 66.40 | 4.55 | 2.90 | B$_{\text{Li}}^{··}$ + 2v$_{\text{Li}}'$ |
| **4a** | 0.5 (B$_2$O$_3$)· 27.5 (Li$_2$O · 2SiO$_2$) | 32.74 | 65.47 | 1.79 | 0.00 | B$_{\text{Si}}'$ + v$_{\text{O}}^{··}$ + v$_{\text{Li}}'$ |
| **4b** | 0.5 (B$_2$O$_3$)· 11.5 (Li$_2$O · 2SiO$_2$) | 31.94 | 63.89 | 4.17 | 0.00 | B$_{\text{Si}}'$ + v$_{\text{O}}^{··}$ + v$_{\text{Li}}'$ |

The incorporated lattice defects in Kröger-Vink notation along with the resulting chemical compositions are summarized in Table 1. Due to the crystal symmetry (space group *Ccc*2),[67] both lattice sites have only one symmetrically non-equivalent position. In all cases, at least three defects have to be incorporated to preserve charge neutrality of the unit cells. It is assumed that the introduced defects are not arbitrarily distributed in the crystal structure, but form defect clusters (Figure 1).

For construction of **1** one Li$^+$ is substituted by Si$^{4+}$ (Si$_{\text{Li}}^{···}$), followed by creation of three Li$^+$ vacancies (v$_{\text{Li}}'$). Since every Li$^+$ is coordinated by five nearby Li$^+$ ions, one with a distance of 2.45 Å and further four 2.95 Å apart, the first v$_{\text{Li}}'$ is located at the closest Li$^+$ position. For the remaining two v$_{\text{Li}}'$ six different initial configurations were constructed to test every combination of v$_{\text{Li}}'$ at the four possible p(v$_{\text{Li}}'$) positions (yellow spheres in Fig. 1). In case of **2**, the SiO$_2$ excess is generated by removing the non-bridging oxygen (v$_{\text{O}}^{··}$) connected to Si$^{4+}$ and creating two Li$^+$ vacancies at the three possible positions p(v$_{\text{Li}}'$) adjacent to v$_{\text{O}}^{··}$, yielding three different initial structures.

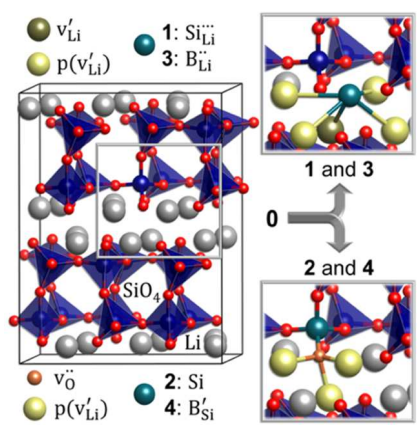

**Figure 1** Incorporation of defect clusters into the lattice of Li$_2$O·2SiO$_2$ (**0**) in vicinity of the Li$^+$ (**1, 3**) and Si$^{4+}$ (**2, 4**) lattice site (green spheres). Different combinations are tested to place Li$^+$ vacancies at the possible positions p(v$_{\text{Li}}'$). Sphere sizes correspond to covalent radii.

Analogous to **1** and **2**, B$^{3+}$ was incorporated at the Li$^+$ and Si$^{4+}$ lattice site for construction of **3** and **4**, respectively. For the substitution of Li$^+$ with B$^{3+}$ (B$_{\text{Li}}^{··}$) in **3**, four initial configurations were constructed, placing one of both v$_{\text{Li}}'$ at the four possible p(v$_{\text{Li}}'$) in the respective structure. These structure modifications also lead to an excess of SiO$_2$ in **3**. For **4** three initial structures were constructed to test every of the three p(v$_{\text{Li}}'$) in vicinity of B$_{\text{Si}}'$.

Relative energies of all models were evaluated with respect to the crystal structures of α-quartz,[68] Li$_2$O[69] and B$_2$O$_3$[70] as the energies $\Delta E_R$ of the reaction:

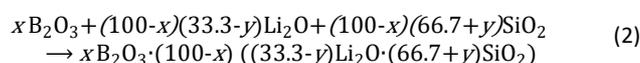

$$x\,\text{B}_2\text{O}_3 + (100\text{-}x)(33.3\text{-}y)\text{Li}_2\text{O} + (100\text{-}x)(66.7\text{+}y)\text{SiO}_2$$
$$\longrightarrow x\,\text{B}_2\text{O}_3\cdot(100\text{-}x)\ ((33.3\text{-}y)\text{Li}_2\text{O}\cdot(66.7\text{+}y)\text{SiO}_2) \qquad (2)$$

## 3 Experimental Procedure

### 3.1 Glass preparation

Glasses with the composition $x$ B$_2$O$_3$·(100-$x$) 33.3 Li$_2$O·66.7 SiO$_2$ ($x$ = 1 and 2 mol%) and (33.3-$y$) Li$_2$O·(66.7+$y$) SiO$_2$ (with $y$ = 1.99, 2.70, 4.74 and 8.67 mol%) were prepared from the raw materials SiO$_2$ (Sipur A1, Bremthaler Quarzizwerke), Li$_2$CO$_3$ (Polskie Odczynniki Chemiczne Gliwice) and H$_3$BO$_3$ (Merck KGaA). Raw materials for 200 g glass were melted in a platinum crucible in the temperature range from 1420 to 1480 °C using an inductively heated furnace for 30 to 60 min. Afterwards the melt was stirred for 90 min using a frequency of 50-60 min$^{-1}$ and subsequently, the melt was kept for further 15 min without stirring. Finally, the melt was cast on a copper and brass block in order to achieve rapid cooling and to avoid crystallization. The glass blocks were then given to a muffle furnace preheated to 465-475 °C which was immediately switched off allowing the glass to cool with a rate of approximately 3 K/min. As stated in Ref. 42, the nucleation rates are not noticeably affected by the cooling procedure and the chemical composition of the glass is not significantly influenced by the melting process, i.e. the lithium evaporation is negligible. Since the prepared glasses are slightly hygroscopic, they were given to a desiccator filled with $P_2O_5$ as drying agent.

**Table 2** Chemical composition of the glasses in mol%, glass transition temperatures measured by DSC and dilatometry, temperatures of the onset of crystallization, $T_{on}$, the maximum of the crystallization peak, $T_c$ and the liquidus temperature, $T_l$.

| Sample name | Chemical composition in mol% | | | Temperature in °C | | | | |
|---|---|---|---|---|---|---|---|---|
| | $Li_2O$ | $SiO_2$ | $B_2O_3$ | $T_g$ | $T_g$(dil) | $T_{on}$ | $T_c$ | $T_l$ |
| A | 33.33 | 66.67 | - | 460 | 464 | 638 | 677 | 1034 |
| B1 | 33 | 66 | 1 | 458 | 468 | 619 | 659 | 1000 |
| B2 | 32.66 | 65.33 | 2 | 458 | 470 | 600 | 642 | 1000 |
| Si1 | 31.34 | 68.66 | - | 460 | 469 | 627 | 659 | 1027 |
| Si2 | 30.63 | 69.37 | - | 455 | 466 | 653 | 682 | 1028 |
| Si3 | 28.59 | 71.41 | - | 461 | 469 | 652 | 694 | 1028 |
| Si4 | 24.66 | 75.34 | - | 465 | - | 681 | 711 | 1033 |

### 3.2 Differential scanning calorimetry

Differential Scanning Calorimetry (DSC) was used to determine the glass transition temperature, $T_g$, as well as the crystallization and melting temperatures using a Linseis DSC PT1600. In order to eliminate sintering effects and to minimize surface crystallization, 60 mg glass powder was remelted in a platinum crucible at 1100 °C, *i.e.*, above the liquidus temperature for 5 min. Then the as prepared sample was cooled in air to room temperature. The subsequent measurements were carried out using a heating rate of 10 K/min.

### 3.3 X-ray diffraction

The glasses and the crystallized samples were studied using X-ray diffraction (XRD). For this purpose, glass samples were thermally treated for 2 h at the temperature of the crystallization peak in the DSC profiles.
The XRD-patterns were recorded using a Rigaku Miniflex 300 diffractometer with Ni-filtered Cu Kα radiation. The scanned 2θ-range was from 10 to 60° and the scan speed was 1°/min. The XRD-patterns were analyzed using the software DIFFRAC.EVA from BRUKER. $Al_2O_3$ was used as internal standard.
Furthermore, crystallized samples were also characterized in a capillary (0.5 mm diameter) in a transmission diffractometer STOE STADI P using Cu Kα radiation.

### 3.4 Viscosity measurements

The viscosities in the range $10^9$ to $10^{12.5}$ dPa·s were determined using a beam bending viscometer Bähr VIS 401. Therefore, glass bars with the dimension 5 x 5 x 50 and 5 x 4 x 50 mm³ were heated with 10 K/min and the bending was measured as a function of the temperature. In order to determine the viscosities in the range $10^{1.3}$ to $10^3$ dPa·s rotation viscometry was applied using a rotation speed of 250 rpm and the cooling rate 5 K/min (Bähr VIS 403). The obtained viscosities $\eta$ as a function of temperature $T$ were

fitted using the Vogel-Fulcher-Tammann (VFT)-equation with the adjustable parameters $C$, $D$ and $T_0$:

$$\eta = C \exp\left(\frac{D}{T - T_0}\right) \tag{3}$$

### 3.5 Determination of the nucleation rates and crystal growth velocities

The nucleation rates and crystal growth velocities were measured in a laser scanning microscope (LSM) AxioImager Z1m PASCAL5 (boron-containing samples) and a LSM 700 (samples with $SiO_2$ excess) which is equipped with a heating stage Linkam HS1500. A detailed explanation of the measurement of the nucleation rates and crystal growth velocities is given in Ref. 42. A stack of several bright field images was recorded along the z-axis of the samples ensuring that the maximum length of the crystal is obtained. Moreover, the number of evaluable crystals could be increased by the use of the z-stack. For the measurement of the crystal growth velocities only crystals which are detached from the others were used. For this reason, the crystal growth velocities were measured in the temperature range 580-660 °C. Otherwise the crystals are not separated and their number is too high. In addition, at the beginning of the measurement the resolution is not sufficient to obtain reliable values and hence higher temperatures are not reasonable.
The activation enthalpy of crystal growth $\Delta H_d$ was determined from an Arrhenius plot of the crystal growth velocity U, *i.e.* log U was plotted as a function of the temperature. The slope of the regression line yields $\Delta H_d$.

## 4 Experimental Results

All analyzed glasses were transparent, homogeneous and do not contain any crystals, bubbles or striae with exception of the samples Si3 and Si4. While the Si3 block shows only some small phase separated areas, the complete Si4 glass block is phase separated. In case of the sample Si3, only glassy areas were used for further analyses.

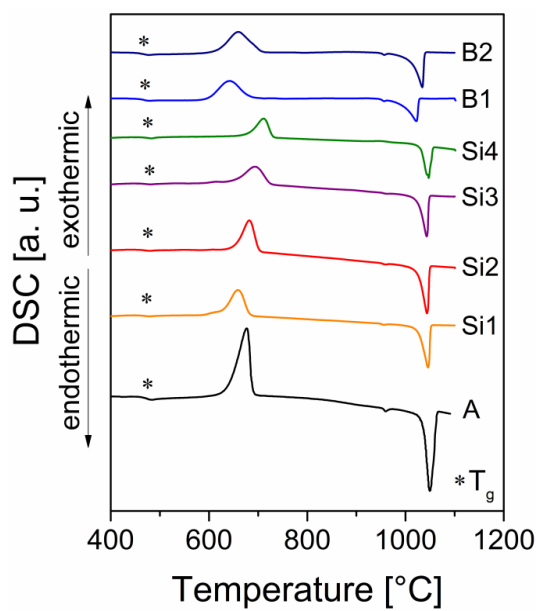

**Figure 2** DSC-profiles of the base glass (Li₂O·2 SiO₂, denoted as A) and the glasses with the compositions $x$ B₂O₃·(100-$x$) 33.3Li₂O·66.7SiO₂ ($x$ = 1 and 2 mol%) and (33.3-$y$) Li₂O·(66.7+$y$)SiO₂ (with $y$ = 1.99 (Si1), 2.70 (Si2), 4.74 (Si3) and 8.67 (Si4) mol%).

Figure 2 shows DSC profiles of the base glass (Li₂O·2 SiO₂, denoted as sample A) and the glasses with the compositions $x$ B₂O₃·(100-$x$) 33.3 Li₂O·66.7 SiO₂ ($x$ = 1 and 2 mol%) and (33.3-$y$) Li₂O·(66.7+$y$) SiO₂ (with $y$ = 1.99 (Si1), 2.70 (Si2), 4.74 (Si3) and 8.67 (Si4) mol%). In all profiles, the glass transition temperature $T_g$ and a distinct crystallization peak are visible. The characteristic temperatures, i.e. the glass transition temperature, $T_g$, the onset of the crystallization peak, $T_{on}$, and the temperature attributed to the crystallization temperature, $T_c$, as well as the liquidus temperature, $T_l$, are summarized in Table 2.

In the B₂O₃ containing glasses, $T_g$ is within the limits of error the same as that of the stoichiometric sample. The values determined by dilatometry are up to 12 K larger than those measured by DSC. The temperatures of the crystallization onset decrease with increasing B₂O₃ additions from 638 to 619 and finally to 600 °C. The same trend is also observed concerning the crystallization peak temperatures. The difference between the crystallization onset and the glass transition temperature, $T_{on}$-$T_g$, which is often used as a measure of glass stability, remains approximately the same. In the case of the samples with higher SiO₂ concentrations, $T_g$ is also within the limits of error the same as for the stoichiometric sample A. While the crystallization onset of the sample Si1 is slightly lower compared with sample A, higher SiO₂ concentrations lead to a significant increase of the crystallization onset to 681 °C and hence to a higher glass stability. Moreover, the addition of B₂O₃ provokes a decrease of the liquidus temperature while the effect of a SiO₂ excess on $T_l$ is only slight. Furthermore, the sample remelted in the DSC crucible and one which was drilled out of the glass block show virtually the same DSC curves (see ESI*, Fig. S1). The small endothermic peaks at about 950 °C were frequently attributed to the transition from the monoclinic to the orthorhombic lithium disilicate phase.[32] However, de Jong et al. only found an orthorhombic lithium disilicate phase and hence this explanation remains controversial.[67]

In Figure 3, the glass viscosities are presented in the temperature range of nucleation (part a) and crystal growth (part b). The increase of the SiO₂ concentration leads to remarkably higher glass viscosities where the effect is more pronounced in the temperature range of crystal growth. While the addition of B₂O₃ causes initially higher viscosities, the viscosities of the samples B1 and B2 are lower than that of the base glass A in the temperature range from 580 to 660 °C. The fitting parameters $C$, $D$ and $T_0$ are listed in Table 3.

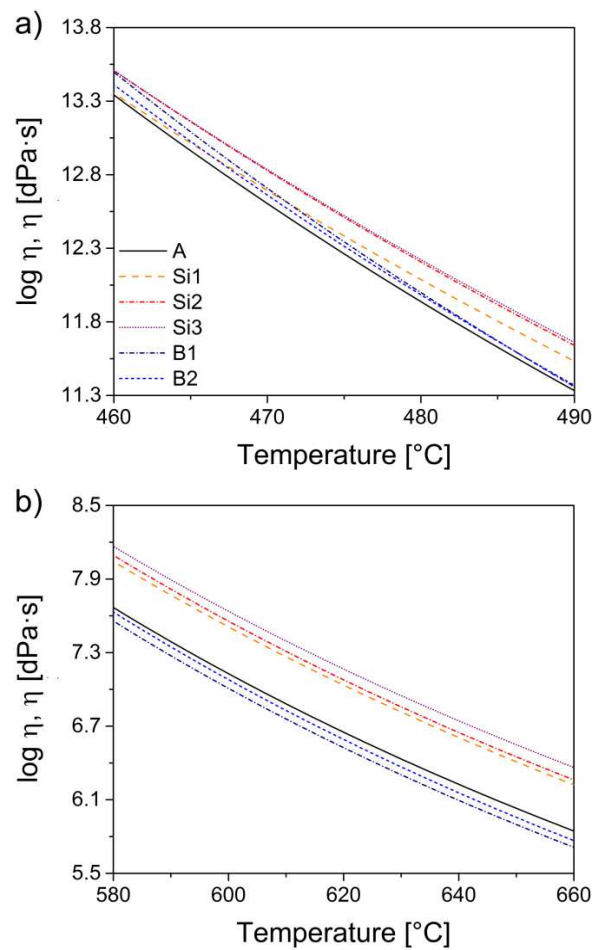

**Figure 3** Viscosities of the glasses determined by fitting the data of the beam bending and rotation viscometry with the VFT-equation (Eq. 3): a) temperature range of nucleation (460-490 °C) and b) studied temperature range of crystal growth (580-660 °C).

**Table 3** Fitting parameters $C$, $D$ and $T_0$. The parameters were determined by fitting the beam-bending and rotation viscosities using the VFT-equation (Eq. 3).

| Sample name | $C$ [dPa·s] | $D$ [°C] | $T_0$ [°C] |
|---|---|---|---|
| A | -1.087 ± 0.013 | 2,641 ± 13 | 277.9 ± 0.8 |
| Si1 | -1.339 ± 0.014 | 3,115 ± 17 | 248.0 ± 1.1 |
| Si2 | -1.210 ± 0.012 | 3,035 ± 14 | 253.8 ± 0.9 |
| Si3 | -0.960 ± 0.009 | 2,966 ± 10 | 255.0 ± 0.6 |

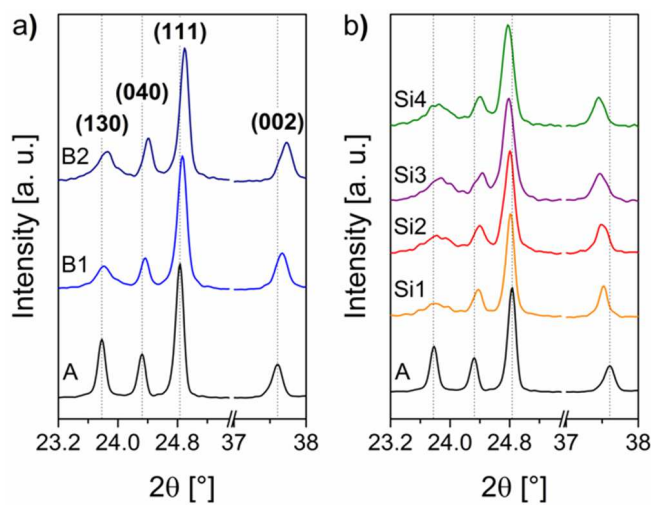

**Figure 4** XRD-patterns of the base glass and the samples doped with $B_2O_3$ (part a) and $SiO_2$ (part b) all crystallized at temperatures of the crystallization peak in the DSC profile.

Figure 4 shows the XRD-patterns of the crystallized samples series $x$ $B_2O_3 \cdot (100-x)$ 33.3 $Li_2O \cdot 66.7$ $SiO_2$ (part a) as well as the series $(33.3-y)$ $Li_2O \cdot (66.7+y)$ $SiO_2$ (part b) in the 2θ-range from 23.2 to 38°. Please note the break at the $x$ axis. All samples show distinct lines attributable to $Li_2O \cdot 2$ $SiO_2$ (JCPDS-no. 82-2396).

According to de Jong et al.,[67] this phase is orthorhombic, belongs to the space group $Ccc2$ and possesses the lattice parameters $a = 5.807$ Å, $b = 14.582$ Å and $c = 4.773$ Å. The addition of $B_2O_3$ leads to a shift of the typical lithium disilicate triplet peaks and the peak at about 2θ = 37.77° to higher 2θ-values (maximum shift 0.12°). On the contrary, an excess of $SiO_2$ only provokes a displacement of the (130) and (040) peaks to higher 2θ-values, while the (111) and (002) peaks, are significantly shifted to lower 2θ-values by up to 0.165°. The lattice parameters of the crystallized lithium disilicate samples were calculated from peak positions obtained by transmission XRD-measurements (see Figure 5 and Table 4). The error of the lattice parameter is about 0.005 Å.

**Table 4** Unit cell parameters of stoichiometric lithium disilicate glass and doped crystallized samples. The values were calculated from peak positions obtained by XRD-measurements using a capillary.

| Sample | Crystallization $T/t$ | Lattice constants in Å | | |
|---|---|---|---|---|
| | | $a$ | $b$ | $c$ |
| A | 677 °C/2 h | 5.820 | 14.625 | 4.781 |
| B1 | 660 °C/2 h | 5.817 | 14.598 | 4.771 |
| B2 | 642 °C/2 h | 5.814 | 14.572 | 4.764 |
| Si1 | 680 °C/2 h | 5.820 | 14.589 | 4.790 |
| Si2 | 680 °C/2 h | 5.817 | 14.580 | 4.793 |
| Si3 | 695 °C/2 h | 5.820 | 14.563 | 4.797 |
| Si4 | 710 °C/2 h | 5.817 | 14.572 | 4.799 |

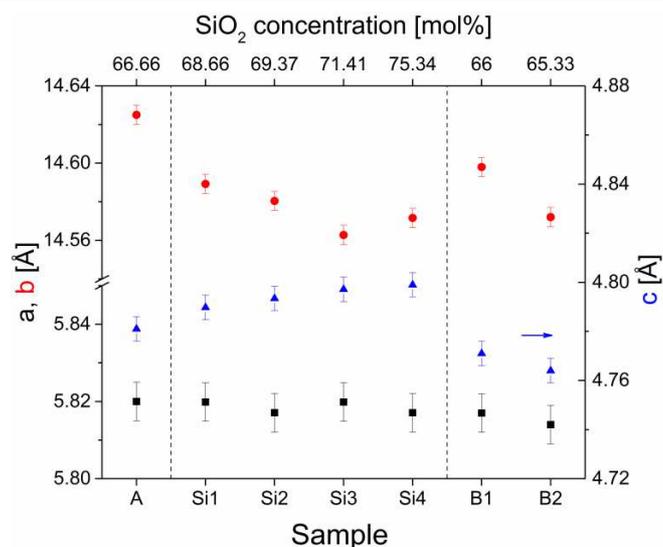

**Figure 5** Change in the lattice parameters as a result of the addition of $B_2O_3$ and the excess of $SiO_2$.

While the stoichiometric lithium disilicate glass shows lattice parameters of $a = 5.820$ Å, $b = 14.625$ Å and $c = 4.781$ Å, the lattice parameters decrease with increasing $B_2O_3$ concentration to $a = 5.814$ Å, $b = 14.572$ and $c = 4.764$ Å. It is seen that the effect is most pronounced for the $b$ axis. In case of the samples with an excess of $SiO_2$, the lattice parameter $b$ decreases while $c$ slightly increases with increasing $SiO_2$ concentration. The variation of the lattice parameter $a$ is within the limits of error. For example, the crystallized Si3 sample shows lattice parameters of $a = 5.820$ Å, $b = 14.563$ Å and $c = 4.797$ Å. The changes in the lattice parameters are a hint at the incorporation of $B_2O_3$ and $SiO_2$ into the $Li_2O \cdot 2$ $SiO_2$ lattice. However, the sites, where boron and silicon are incorporated cannot be concluded from the XRD-patterns.

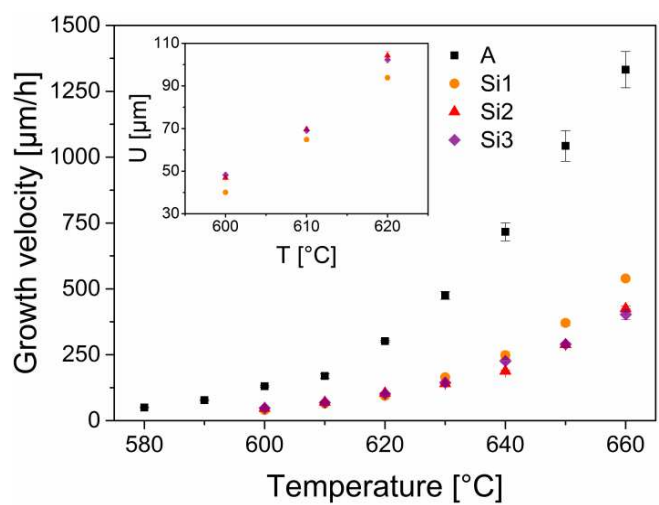

**Figure 6** Crystal growth velocities as a function of the temperature: effect of the $SiO_2$ excess.

In Figure 6, the crystal growth velocities of samples in the series $(33.3-y)$ $Li_2O\cdot(66.7+y)$ $SiO_2$ are displayed as a function of the temperature. The values were obtained by in situ measurements in the heating stage of a laser scanning microscope. In all analyzed samples the lithium disilicate crystals possess an elliptical morphology. As shown in Figure 6, the increase of the $SiO_2$ concentration causes a drastic decrease of the crystal growth velocities. In the inset, the crystal growth velocities in the temperature range from 600 to 620 °C are shown in more detail. It is apparent that only slight deviations between the samples, especially between Si2 and Si3, occur. Comparing the crystal growth velocities of the latter samples (Table 5), it becomes apparent that the growth velocity of the lithium disilicate crystals is almost equal in both samples despite the considerably higher $SiO_2$ concentration of sample Si3. The addition of $B_2O_3$ also leads to a reduction of the crystal growth velocities; however, the effect is not as pronounced as in the case of the samples with an excess of $SiO_2$. Moreover, the activation enthalpy for crystal growth $\Delta H_d$ was determined from Arrhenius plots of the crystal growth velocity. While $\Delta H_d$ of the $B_2O_3$ containing samples is almost unchanged, the sample Si1 shows a slightly higher $\Delta H_d$ value. However, a further increase of the $SiO_2$ concentration provokes a remarkable decrease of the activation enthalpy.

# 5 Computational results

Table 6 shows the comparison of the experimentally determined (A) and DFT calculated (0) cell parameters $a$, $b$, $c$ of $Li_2O\cdot2SiO_2$ using different methodologies. The PBEsol exchange-correlation functional augmented by the dispersion correction (D3) provides the best agreement with experimental values with the largest absolute deviation of -0.04 Å for $a$ and less than -0.01 Å for $b$.

**Table 5** Crystal growth velocities $U$ as a function of the temperature and activation enthalpies for crystal growth $\Delta H_d$ [kJ/mol].

| $T$ [°C] | $U$ [µm/h] | | | | | |
|---|---|---|---|---|---|---|
| | A[a] | B1[b] | B2[b] | Si1 | Si2 | Si3 |
| 580 | 50 ±1 | - | - | - | - | - |
| 590 | 78 ±3 | - | - | - | - | - |
| 600 | 131 ±3 | 74 ±1 | 68 ±1 | 40 ±1 | 47 ±1 | 48 ±1 |
| 610 | 169 ±9 | 122 ±1 | 111 ±4 | 65 ±1 | 70 ±1 | 69 ±1 |
| 620 | 302 ±4 | 179 ±1 | 184 ±4 | 94 ±1 | 104 ±2 | 102 ±1 |
| 630 | 475 ±15 | 309 ±8 | 286 ±2 | 165 ±2 | 140 ±2 | 144 ±4 |
| 640 | 717 ±34 | 458 ±6 | 422 ±5 | 248 ±13 | 188 ±4 | 227 ±6 |
| 650 | 1,043 ±58 | 630 ±4 | 621 ±5 | 370 ±12 | 289 ±6 | 290 ±4 |
| 660 | 1,332 ±69 | 902 ±9 | 981 ±28 | 539 ±9 | 425 ±10 | 404 ±20 |
| $\Delta H_d$ | 283 ±4 | 280 ±8 | 288 ±8 | 296 ±5 | 243 ±5 | 241 ±6 |

[a] values taken from Ref. 42, [b] values taken from Ref. 48

**Table 6** Deviation of the calculated cell parameters $a$, $b$, $c$ [Å] of $Li_2O\cdot2SiO_2$ from the experimental value for A.

| | $a$ | $b$ | $c$ |
|---|---|---|---|
| A (exp) | 5.820 | 14.625 | 4.781 |
| PBE | +0.052 | +0.142 | +0.063 |
| PBE+D3 | -0.015 | +0.060 | +0.028 |
| PBEsol | +0.009 | +0.066 | +0.011 |
| PBEsol+D3 | -0.041 | -0.008 | -0.013 |

Figure 7 shows the cell parameter $b$ of $x$ $B_2O_3\cdot(100\text{-}x)$ $((33.3\text{-}y)Li_2O\cdot(66.7+y)SiO_2)$ as a function of the chemical composition $(x, y)$ obtained from experiments and DFT calculations for the most stable structures of **0** up to **4**. The trend of a decreasing cell parameter $b$ with increasing $B_2O_3$ content $x$ as well as $SiO_2$ excess $y$ observed in experiment is in a very good agreement with DFT calculations for the incorporation of $Si^{4+}$ and $B^{3+}$ on the $Li^+$ lattice site (**1** and **3**). This is not the case for the introduction of defects in vicinity of the $Si^{4+}$ lattice site (**2** and **4**).

The optimized unit cell of **0b** and the energetically most stable structures obtained for **1b** up to **4b** are depicted in Figure 8. The moieties [$SiO_4$], [$Si_2O_6$], [$BO_4$] and [$BO_3$] shown in Fig. 8 were also found for the most stable structures of the remaining unit cell sizes (cf. Table 1). Comparison of calculated cell parameters and reaction energies $\Delta E_R$ (cf. Eq. 2) of **0** with the lowest energy configurations for **1** up to **4** is presented in Table 7. In addition, results for the second most stable structure of **3b**, denoted as **3b\***, are shown.

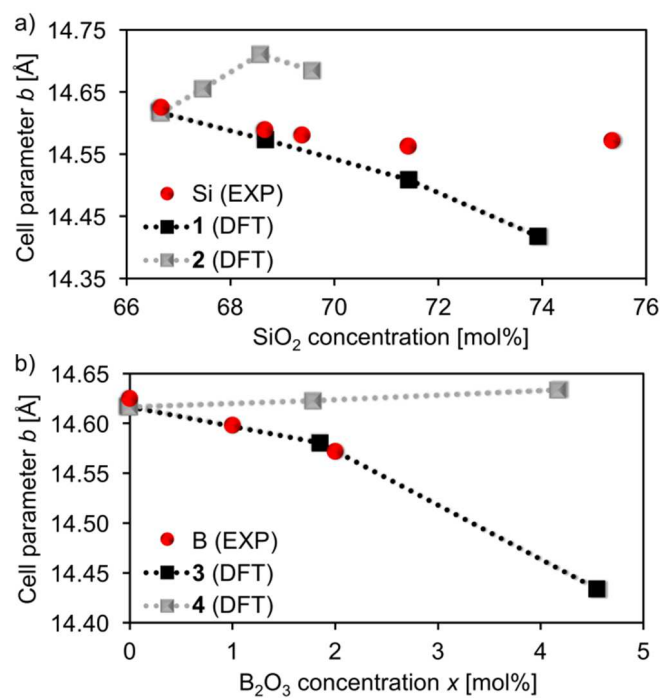

**Figure 7** Calculated and experimental cell parameter $b$ for $x$ $B_2O_3\cdot(100\text{-}x)$ $((33.3\text{-}y)Li_2O\cdot(66.7+y)SiO_2)$ as a function of a) $SiO_2$ concentration and b) $B_2O_3$ concentration $x$.

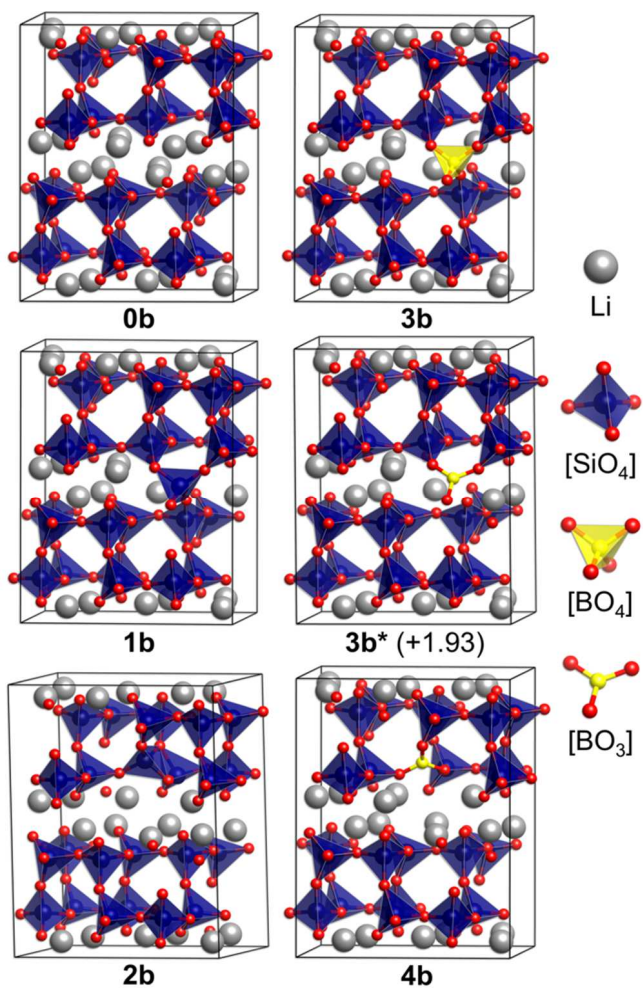

**Figure 8** Optimized crystal structure of Li₂O·2SiO₂ (**0b**) as well as unit cells with Si excess (**1b**, **2b**) and B substitution (**3b**, **4b**). Relative energy of **3b\*** with respect to **3b** is given in parentheses [kJ/mol].

The unit cell of **0** contains two silica sheets consisting of [SiO₄] tetrahedra which are separated by a layer of Li⁺ ions. With substitution of an additional Si⁴⁺ ion at the Li⁺ lattice site (**1**), a new [SiO₄] tetrahedron within the Li⁺ layer is formed connecting both silica sheets. This leads to a contraction of the crystallographic *b* axis. In case of **2** two [SiO₄] tetrahedra within the silica sheet are rearranged resulting in an edge-sharing [Si₂O₆] unit and only a slight change of cell parameters. However, **2** is considerably higher in energy compared to **1**.

Similar to **1**, the incorporation of B³⁺ in the Li⁺ lattice site yields a [BO₄] tetrahedron within the Li⁺ layer, which is also accompanied by contraction of the *b* axis. The second most stable structures **3a\*** and **3b\***, which are both at most 2.5 kJ/mol higher in energy, show three-fold coordinated [BO₃] that is also bonded to both silica sheets. In these cases, the reduction of the cell parameter *b* is less pronounced compared to **3a** and **3b**. When Si⁴⁺ is substituted by B³⁺ (**4**), trigonal [BO₃] units are formed within the silica sheet leaving the cell parameters virtually unaffected. Compared to **3a** and **3b**, both **4a** and **4b** are slightly more stable with energy differences of only 0.4 and 3.7 kJ/mol, respectively.

In general, the formation reactions for all models with respect to the crystalline unary oxides (SiO₂, Li₂O, B₂O₃) according to Eq. 2 are considerably exothermic. However, all defect containing configurations are thermodynamically less stable than pure Li₂O·2SiO₂ (**0**). In addition, $\Delta E_R$ increases with higher dopant concentrations. Independent of the chemical composition and the incorporated lattice defects, the cell parameters *a* and *c* remained virtually constant with changes of at most 0.06 Å for all models. In contrast, the crystallographic *b* axis shows considerable contractions for **1** and **3** ranging from 0.04 to 0.20 Å as well as moderate expansions between 0.01 and 0.1 Å for **2** and **4**.

## 6 Discussion

The systematic shift of the observed XRD-patterns and lattice parameters with varying chemical composition (Fig. 4, Table 4) suggests that B₂O₃ and SiO₂ are incorporated into the lattice of Li₂O·2SiO₂. This is clearly supported by the remarkable agreement of the systematic decrease of cell parameter *b* obtained from experiments and DFT calculations for the incorporation of Si⁴⁺ and B³⁺ on the Li⁺ lattice site of Li₂O·2SiO₂ (**1** and **3**, Fig. 7). Results of XRD measurements show for the crystallographic *a* and *c* axes only minor changes of at most 0.02 Å for all investigated compositions. These observations coincide with DFT simulations showing cell parameter changes of less than 0.06 and 0.04 Å for the incorporation of SiO₂ and B₂O₃, respectively. Such small variations of the *a* and *c* axes lie in the same order of magnitude as the accuracy of the employed simulation method (*cf*. Tables 6 and 7). Therefore, both cell parameters can be assumed to be unaffected by the addition the SiO₂ and B₂O₃ within the investigated composition ranges.

**Table 7** Lattice parameters of Li₂O·2SiO₂ (**0**) and changes in the lattice parameters *a*, *b*, *c* [Å] caused by the incorporation of lattice defects (**1 - 4**). Reaction energies $\Delta E_R$ [kJ/mol] (*cf*. Eq. 2).

|        | *a*     | *b*     | *c*     | $\Delta E_R$ |
|--------|---------|---------|---------|---------|
| **0a** | 15.427  | 14.617  | 12.503  | -123.9  |
| **1a** | 0.000   | -0.043  | +0.005  | -118.6  |
| **2a** | -0.038  | +0.039  | +0.013  | -106.6  |
| **3a** | +0.002  | -0.079  | -0.020  | -115.9  |
| **3a\*** | +0.038 | -0.036  | -0.027  | -113.5  |
| **4a** | -0.025  | +0.006  | +0.006  | -116.4  |
| **0b** | 7.492   | 14.617  | 11.150  | -123.9  |
| **1b** | +0.002  | -0.108  | +0.008  | -109.8  |
| **2b** | -0.056  | +0.094  | +0.016  | -85.5   |
| **3b** | -0.021  | -0.183  | +0.003  | -103.1  |
| **3b\*** | +0.015 | -0.128  | -0.019  | -101.2  |
| **4b** | +0.010  | +0.017  | -0.043  | -106.9  |
| **0c** | 5.779   | 14.617  | 9.536   | -123.9  |
| **1c** | +0.035  | -0.199  | +0.064  | -104.9  |
| **2c** | +0.012  | +0.068  | -0.036  | -64.3   |

For the crystallization of $x$ $B_2O_3\cdot(100-x)$ (($33.3-y$)$Li_2O\cdot(66.7+y)SiO_2$) glasses containing excessive $SiO_2$, the experimentally observed cell parameter $b$ remains approximately constant above 15 mol% $SiO_2$ excess. By contrast, **1** shows a pronounced contraction of the $b$ axis also at larger $SiO_2$ concentrations (>71.5 mol% $SiO_2$). This indicates that $SiO_2$ is only partially incorporated into the $Li_2O\cdot2SiO_2$ lattice during the crystallization of glasses with $SiO_2$ concentrations larger than 71.5 mol%. Similarly, it was concluded in a previous study of Glasser[32] that ($33.3-y$)$Li_2O\cdot(66.7+y)SiO_2$ solid solutions are only formed up to $SiO_2$ concentrations of 72.3 mol% $SiO_2$.

In contrast to **1** and **2** that are energetically well separated (Table 7), DFT simulations revealed that $B^{3+}$ can be incorporated into the $Li_2O\cdot2SiO_2$ lattice in form of three possible configurations showing similar reaction energies $\Delta E_R$ (Eq. 2): as [$BO_4$]- (**3**) or [$BO_3$]-unit (**3\***) on the $Li^+$ lattice site as well as [$BO_3$]-unit on the $Si^{4+}$ lattice site (**4**). The latter is only 0.4 and 3.7 kJ/mol more stable compared to **3** for $B_2O_3$ concentrations of about 2.0 and 4.5 mol%, respectively. Due to such small energy differences evaluated at $T = 0$ K, $i.e.$, without consideration of vibrational contributions to the lattice free energy, it cannot be eventually clarified which configuration is the thermodynamically most stable one at the crystallization temperature. However, **4** has virtually the same lattice parameters compared with $Li_2O\cdot2SiO_2$ (**0**), whereas structure optimizations for **3** yielded a contraction of the $b$ axis in very good agreement with experiments (Fig. 7b). This indicates that **3** is the preferred configuration evolving from crystallization, without ruling out the formation of **3\*** and **4**. Since **3** shows also a higher $SiO_2$ concentration compared to stoichiometric $Li_2O\cdot2SiO_2$ ($cf.$ Table 1), the crystallization of **3** would change the chemical composition at the crystallization front, if the parent glass has not the corresponding Si/Li ratio. In such a case, this might have a crucial influence on the glass viscosities and crystallization kinetics.

The addition of minor concentrations of $B_2O_3$ leads to a slight increase of the viscosities in the temperature range of nucleation, but at higher temperatures when crystal growth takes place, the viscosities of the boron containing samples are somewhat lower than those of the undoped glass. Generally, $B_2O_3$ is incorporated in the glass network as network modifier and may occur as [$BO_4$]-tetrahedra or [$BO_3$]-units. In a glass with high network modifier concentrations as in the studied compositions, predominantly [$BO_4$]-tetrahedra are formed at not too high temperatures as in the case of the supplied nucleation temperatures.[71] As stated in Ref. 48, at higher temperatures the equilibrium between the [$BO_4$]-tetrahedra and the [$BO_3$]-units should be shifted towards [$BO_3$]-units. Then, the number of non-bridging oxygens increases which easily explains the lower viscosities observed using rotation viscometry.[72,73] On the contrary, the glasses with an excess in $SiO_2$ show higher viscosities in the entire temperature range which can be ascribed to the lower concentration of the network modifier $Li_2O$ causing a higher network connectivity and hence a viscosity increase. A calculation of the viscosities using for example the Flügel model[74] was not possible since the compositions are outside the composition/interaction limits and hence the viscosity predictions are expected to have a too large error. However, the model "Priven 2000" ($e.g.$ Refs. 75,76) in the software SciGlass can be used for these glass compositions, but a prediction using this model leads to values which remarkably differ from the viscosities obtained by beam bending (see ESI, Fig. S2). For this reason, this model is also not appropriate to predict viscosities of the glass compositions investigated in this work.

The reaction energies $\Delta E_R$ (Eq. 2) for the formation of $x$ $B_2O_3\cdot(100-x)$ (($33.3-y$)$Li_2O\cdot(66.7+y)SiO_2$) solid solutions with respect to the unary oxides ($SiO_2$, $Li_2O$, $B_2O_3$) are negative for all optimized unit cells. The value calculated for pure $Li_2O\cdot2SiO_2$ (**0**) of -124 kJ/mol is in a good agreement with the experimentally determined reaction enthalpy of (-140 ± 3) kJ/mol.[77] However, each crystal structure incorporating of $SiO_2$ and $B_2O_3$ is energetically less stable than **0** due to the introduction of additional vacancies, $v'_{Li}$ and $v^{\cdot\cdot}_O$. For the vacancy incorporation into **0**, it was assumed that all lattice defects are clustering around the substitutional ion. Indeed, previous computational studies demonstrated that the association of such defects is favored in oxide materials, if the ionic radius of the dopant is lower in comparison to that of the host ion.[50,51] This is the case for all substitutions investigated in this work, namely $Si^{\cdot\cdot\cdot}_{Li}$, $B^{\cdot\cdot}_{Li}$ and $B'_{Si}$. Moreover, a strong binding between $M^{\cdot}_{Li}$ and $v'_{Li}$ was found for Li$M$PO$_4$ ($M$ = Fe$^{2+}$, Mn$^{2+}$, Co$^{2+}$, Ni$^{2+}$)[49] that also supports the assumption of preferential formation of defect clusters in ionic Li compounds.

The increase of $\Delta E_R$ with higher dopant and, consequently, larger defect concentrations provides a further verification $x$ $B_2O_3\cdot(100-x)$ (($33.3-y$)$Li_2O\cdot(66.7+y)SiO_2$) solid solutions are metastable with respect to pure $Li_2O\cdot2SiO_2$. Therefore, the formation of these solid solutions as well as the quantity of dopants incorporated into the host lattice strongly depends on the crystallization kinetics that is determined by the processes taking place at the crystallization front. Since the crystal growth velocities of the glasses decrease with increasing $B_2O_3$[48] and $SiO_2$ concentration, it is assumed that in course of the crystallization, the chemical composition of the residual glass matrix changes near the crystallization front hindering nucleation and crystal growth.

As indicated by the comparison of the observed and calculated cell parameter $b$ for the systems containing excessive $SiO_2$ (Fig. 7a), $SiO_2$ is probably only partially incorporated into $Li_2O\cdot2SiO_2$ leading to an enrichment of $SiO_2$ at the crystallization front. This increases the viscosity and, consequently, decreases the diffusivity at the interface between precipitated crystal and glass matrix yielding a decrease of the crystal growth velocity. Obviously, the crystal growth velocities of the samples Si2 and Si3 are almost the same which might be attributed to the fact that with higher $SiO_2$ excess also the quantity of incorporated $Si^{4+}$ increases. As already mentioned, Glasser[32] observed that the formation of the ($33.3-y$)$Li_2O\cdot(66.7+y)SiO_2$ solid solutions extends to glasses which contain 66.66 to 72.3 mol% $SiO_2$ and hence the glass compositions Si1, Si2 and Si3 lie in this range. Unfortunately, the hypothesis of Glasser cannot be verified by sample Si4 as the glass was phase separated after the preparation and hence a determination of the crystal growth velocities was not possible.

In the case of the $B_2O_3$ containing glasses, the very good agreement of the calculated and experimentally determined cell parameter $b$ (Fig. 7b) suggests that the $B_2O_3$ concentration in the precipitated crystals is similar to the parent glass compositions, at least for crystals formed in the early stages of nucleation and crystal growth. With proceeding crystal growth, the formation of **3** would lead to an enrichment of $Li_2O$ (or depletion of $SiO_2$) in the surrounding glass matrix that shifts the chemical equilibrium in such a manner that a further formation of **3** is less favorable. Consequently, during the further course of crystallization, $B^{3+}$ is either incorporated on the $Si^{4+}$ lattice sites (**4**) or could be more and more accumulated at the crystallization front. In the latter case, the enrichment of $[BO_3]$- and $[BO_4]$-units at the interface between the crystalline and amorphous phase could interfere the attachment of building blocks required for further crystal growth, *i.e.*, lowering the so-called impingement rate. This might rationalize the decrease of the crystal growth velocities observed for B1 and B2 despite their lower viscosities compared to A.

In addition, the accumulation of $Li^+$ in the residual glass matrix caused by the formation of **3** could yield $(33.3-y)Li_2O\cdot(66.7+y)SiO_2$ solid solutions showing an excess of $Li_2O$ ($y < 0$), which might also influence the $B^{3+}$ incorporation and, hence, the crystal growth velocities in later stages of crystallization. These $Li_2O$-rich solid solutions extend in the $Li_2O$-$SiO_2$ phase diagram from $SiO_2$ concentrations of 66.7 to about 62 mol%.[78–80] Indeed, the crystallization of solid solutions showing a time dependent change of their chemical composition and crystal growth rates was reported for $Na_2O\cdot2CaO\cdot3SiO_2$ and $CaO\cdot Al_2O_3\cdot SiO_2$ glasses.[81–83] Moreover, numerous studies discussed that the crystallization of stoichiometric $Li_2O\cdot2SiO_2$ and its $(33.3-y)Li_2O\cdot(66.7+y)SiO_2$ solid solutions could involve the formation of metastable intermediate phases, among them lithium metasilicate $Li_2O\cdot SiO_2$.[15,39,80,84] Accordingly, the addition of $B_2O_3$ has very probably a decisive influence on both, the compositional change of the glass matrix in the course of crystallization and the thermodynamic stability of $(33.3-y)Li_2O\cdot(66.7+y)SiO_2$ solid solutions including possible intermediate phases. Therefore, achieving a deeper understanding of these crystallization mechanisms requires further computational and experimental studies on $B_2O_3$ containing $(33.3-y)Li_2O\cdot(66.7+y)SiO_2$ solid solutions having also an excess of $SiO_2$ ($y > 0$) and $Li_2O$ ($y < 0$), respectively.

## Conclusions

In summary, DFT simulations in combination with experimental investigations on the crystallization of $SiO_2$ and $B_2O_3$ doped glasses provides the first clear indication for the formation of metastable $x$ $B_2O_3\cdot(100-x)$ $((33.3-y)Li_2O\cdot(66.7+y)SiO_2)$ solid solutions. DFT calculations revealed that both $Si^{4+}$ and $B^{3+}$ are preferentially incorporated into the

$Li_2O\cdot2SiO_2$ crystal structure by the substitution of $Li^+$ resulting in a considerable contraction of the crystallographic $b$ axis, which is in a very good agreement with experimentally determined lattice parameters. Since the incorporation of $SiO_2$ and $B_2O_3$ is connected with the creation of $Li^+$ vacancies, the formed solid solutions are energetically less stable compared to pure $Li_2O\cdot2SiO_2$. Therefore, their formation process and, consequently, the quantity of the incorporated dopants is determined by the crystallization kinetics. By doping the glasses with additional $SiO_2$ ($y > 0$), an increase of the glass viscosities was observed leading to a decrease of the crystal growth velocities, since excessive $SiO_2$ is only partially incorporated into the $Li_2O\cdot2SiO_2$ lattice. By contrast, the addition of $B_2O_3$ ($x > 0$) results in a decrease of both the glass viscosity and crystal growth velocity. This could be attributed to an enrichment of $Li^+$ in the residual glass matrix, due to the creation of an excess of $SiO_2$ in the precipitated crystal phase when $B^{3+}$ is incorporated (either as $[BO_4]$- or $[BO_3]$-unit) on the $Li^+$ lattice site. Consequently, the affinity for further $B^{3+}$ incorporation is reduced resulting in an accumulation of $[BO_4]$- or $[BO_3]$-units at the crystallization front that might interfere the crystal growth due to a lowered impingement rate. In addition, the formation of $(33.3-y)Li_2O\cdot(66.7+y)SiO_2$ solid solutions enriched with $Li_2O$ ($y < 0$) could have a crucial influence upon $B_2O_3$ incorporation and, therefore, on the crystal growth velocities.

## Conflict of interest

There are no conflicts to declare.

## Acknowledgements

Authors gratefully acknowledge the financial support from the German Science Foundation (DFG), Fonds der Chemischen Industrie and Turbomole GmbH.